\documentclass[prd,twocolumn,showpacs,superscriptaddress,footinbib]{revtex4-1}
\usepackage{amsfonts}
\usepackage{amsmath}
\usepackage{amssymb}
\usepackage{bm}
\usepackage{dcolumn}
\usepackage{graphicx}
\usepackage{graphics}
\usepackage{latexsym}
\usepackage{rotating}
\usepackage[colorlinks=true]{hyperref}
\usepackage[all]{hypcap} 
\usepackage{xspace} 
\usepackage[usenames]{xcolor}
\usepackage{mathrsfs}
\usepackage{multirow}

\usepackage{ulem}
\definecolor {darkgreen}{rgb}{0.2,0.7,0.2}

\newcommand\be{\begin{equation}}
\newcommand\ee{\end{equation}}
\newcommand\bw{\begin{widetext}}
\newcommand\ew{\end{widetext}}

\newcommand{\bea}{\begin{eqnarray}}
\newcommand{\eea}{\end{eqnarray}}

\newcommand{\ep}{\varepsilon}

\def\apj{Astrophys. J. \ }%
\def\prc{Phys. Rev. C\ }%
\def\prd{Phys. Rev. D \ }%
\def\nat{Nature}%
\begin{document}
\title{ Implications from GW170817 and I-Love-Q relations for
  relativistic hybrid stars }

\author{Vasileios Paschalidis} \affiliation{Theoretical Astrophysics
  Program, Departments of Astronomy and Physics, University of
  Arizona, 933 N. Cherry Ave., Tucson, AZ 85721}

\author{Kent Yagi}
\affiliation{Department of Physics, University of Virginia, Charlottesville, Virginia 22904, USA}

\author{David Alvarez-Castillo}
\affiliation{Bogoliubov Laboratory of Theoretical Physics, Joint Institute for Nuclear Research,
	141980 Dubna, Russia}
\affiliation{ExtreMe Matter Institute EMMI, GSI Helmholtzzentrum f\"ur Schwerionenforschung, Planckstra\ss e 1, 64291 Darmstadt, Germany}

\author{David B.~Blaschke}
\affiliation{Bogoliubov Laboratory of Theoretical Physics, Joint Institute for Nuclear Research, 141980 Dubna, Russia}
\affiliation{Institute of Theoretical Physics, University of Wroclaw, 50-204 Wroclaw, Poland}
\affiliation{National Research Nuclear University (MEPhI), 115409 Moscow, Russia}

\author{Armen Sedrakian}
\affiliation{Frankfurt Institute for Advanced Studies, D-60438 
  Frankfurt-Main, Germany }

\begin{abstract}
Gravitational wave observations of GW170817 placed bounds on the tidal
deformabilities of compact stars allowing one to probe equations of
state for matter at supranuclear densities. Here we design new
parametrizations for hybrid hadron-quark equations of state, that
give rise to low-mass twin stars, and test them against GW170817. We
find that GW170817 is consistent with the coalescence of a binary
hybrid star--neutron star. We also test and find that the I-Love-Q
relations for hybrid stars in the third family agree with those for
purely hadronic and quark stars within $\sim 3\%$ for both slowly and
rapidly rotating configurations, implying that these relations can be
used to perform equation-of-state independent tests of general
relativity and to break degeneracies in gravitational waveforms for
hybrid stars in the third family as well.
\end{abstract}

\date{\today}
\maketitle


\section{Introduction}

The direct detection of gravitational waves
(GWs) by the LIGO and Virgo Scientific Collaborations (LVC) has
already started to revolutionize our understanding of the cosmos. The
LVC direct detections of GWs, consistent with the inspiral and merger
of binary black
holes~\cite{FirstDirectGW,Abbott:2016nmj,Abbott:2017vtc,Abbott:2017oio,Abbott:2017gyy},
solidified the onset of the era of GW astronomy, and provided a number
of astrophysics and fundamental physics implications (see,
e.g.,~\cite{TheLIGOScientific:2016htt,TheLIGOScientific:2016src,Yunes:2016jcc,Abbott:2017vtc}). The
recent simultaneous detection of a GW signal by the
LVC~\cite{TheLIGOScientific:2017qsa} (event GW170817) and a gamma-ray
burst (GRB) by the Fermi satellite~\cite{GW170817_GRB170817},
including subsequent counterpart electromagnetic
signals~\cite{multimessGW170817,Abbott:2017wuw,Smartt:2017fuw,Cowperthwaite:2017dyu,Nicholl:2017ahq,Chornock:2017sdf,Margutti:2017cjl,2017Sci...358.1556C},
have ushered us in the era of ``multimessenger'' astronomy,
astrophysics and cosmology~\cite{Abbott:2017xzu}.

The impact of GW170817 on astrophysics~\cite{Metzger:2017wot} and
fundamental physics is far reaching. For example, GW170817 and GRB
170817A provided the best evidence, yet, that some GRBs are associated
with the merger of binary compact stars as envisioned
by~\cite{Pacz86,EiLiPiSc,NaPaPi} and recently demonstrated numerically
in~\cite{prs15,Ruiz:2016rai} (see
also~\cite{Paschalidis:2016agf,Baiotti:2016qnr} for recent
reviews). GW170817 placed constraints on the properties of the
progenitor of the binary compact object
GW170817~\cite{Abbott:2017ntl}, and the GW background from compact
binaries~\cite{Abbott:2017xzg}. Moreover, GW170817 and GRB 170817A
constrained the speed of gravity, the equivalence principle and
Lorentz invariance~\cite{GW170817_GRB170817}, consequently
constraining to a large degree gravity theories designed to explain
the accelerated expansion of the Universe without dark
energy~\cite{Sakstein:2017xjx,Jana:2017ost,Nojiri:2017hai,Green:2017qcv,Baker:2017hug,Ezquiaga:2017ekz,Creminelli:2017sry,Heisenberg:2017qka}
(see also~\cite{Lombriser:2015sxa,Lombriser:2016yzn} for earlier
work). Furthermore, GW170817 furnished the first ever ``standard
siren''~\cite{Schutz:1986gp} measurement of the Hubble
constant~\cite{Abbott:2017xzu}.

Another important impact of GW170817 on fundamental physics is that
GW170817 set bounds on the tidal deformabilities (TDs) of compact
stars~\cite{TheLIGOScientific:2017qsa}. The observation of $2M_\odot$
pulsars~\cite{Demorest2010,Antoniadis2013,Fonseca:2016tux,Arzoumanian:2017puf} had already
set a tight constraint on the properties of nuclear matter, requiring
its equation of state (EOS) to be stiff, see,
e.g.,~\cite{BeniBlaschke2015A&A...577A..40B,Alvarez-Castillo:2017qki,AlfordSedrakian17}. However,
GW170817 has ``raised the bar'' compact star EOSs must pass to be
physically viable: {\it candidate EOSs must now also satisfy the
  GW170817 constraints on the TD of compact stars.}

A study of the consequences of the GW170817 TD bounds on the nuclear
EOS was performed in~\cite{Annala:2017llu}. Multimessenger
observations of GW170817 were also used to place constraints on
nonspinning neutron star (NS) masses and radii using approaches with
a varying number of assumptions, see,
e.g.,~\cite{Bauswein:2017vtn,Ruiz:2017due,Radice:2017lry,Rezzolla:2017aly,Shibata:2017xdx,Margalit:2017dij,Zhou:2017pha}. However,
these previous works did not consider EOSs that support a strong phase
transition with a sufficiently large jump in energy density to give
rise to a separate (third family) branch of compact stars like the
EOSs we develop here.

In this work, we investigate how GW170817 can constrain the properties
of {\emph{hybrid compact stars}} which have a strong hadron-quark
phase transition in their interiors. In particular, we mainly focus on
EOSs allowing a \emph{third family} of stable compact objects at low
mass, i.e., in addition to the stable branches of white dwarfs and
NSs. The third family of compact stars, which has been studied over
several
decades~\cite{Gerlach1968PhRv..172.1325G,Kampfer1981JPhA...14L.471K,Schertler2000NuPhA.677..463S,Glendenning2000A&A...353L...9G}),
arises when there is an instability region separating hadronic NSs
from hybrid stars (HSs); this leads to the emergence of twins - NSs
and HSs having the same mass but different
radii~\cite{Glendenning2000A&A...353L...9G}.  The HS internal
structure requires a single-phase quark core enclosed by a hadronic
shell with a first-order phase transition at their interface. An
additional phase transition in the quark core can lead to a fourth
family of compact stars~\cite{AlfordSedrakian17}, but we do not
consider this possibility here.

In this paper we develop \emph{new parametrizations} of hybrid
hadron-quark EOSs that allow for a third family of compact stars to
emerge at ``low mass'' ($\sim 1.5M_\odot$) \emph{and} are consistent
with the existence of $2M_\odot$ pulsars. We investigate whether these
EOSs are consistent with GW170817 by computing the TD of corresponding
compact stars. Moreover, we compute the I-Love-Q
relations~\cite{Yagi:2013bca,Yagi:2013awao} (see
also~\cite{Yagi:2016bkt,Paschalidis:2016vmz,Doneva:2017jop} for recent
reviews) for slowly and rapidly rotating HSs in the third family
constructed with these EOSs and compare them to the I-Love-Q relations
of purely hardronic and quark stars. It is important to test the
universality of I-Love-Q relations because HSs in the third family
have a sharp first-order phase transition at the hadron-quark
interface in their interior, and it is not a priori clear that such
HSs satisfy the neutron and quark star I-Love-Q
relations~\footnote{Note that in~\cite{Bandyopadhyay:2017dvi}, which
  appeared after our work, it is claimed that the I-Love-Q relations
  do not hold for HSs}. Once established also for 3rd family and
hybrid stars, the I-Love-Q relations can be used to perform
equation-of-state independent tests of general relativity and to break
degeneracies in GWs~\cite{Yagi:2013bca,Yagi:2013awao}. Thus, it is
important to know if these relations hold for HSs in the third family.
Throughout, we adopt geometrized units unless otherwise stated.

\section{Equations of State}

The new parametrizations of EOSs we
develop here describe zero-temperature nuclear matter in
$\beta$-equilibrium with a low-density phase of nucleonic matter and
high-density phase of quark matter. We consider two sets of EOSs which
cover a range of current models as described
in~\cite{AlfordSedrakian17} (Set-I) and
\cite{Alvarez-Castillo:2017qki} (Set-II).

In Set I the low-density phase is based on a covariant density
functional theory~\cite{RingReview2010} with density-dependent
couplings~\cite{DDME2005}, as applied to hadronic matter
in~\cite{ColucciSedrakian13}.  The Lagrangian underlying the density
functional, and the corresponding zero-temperature pressure of
nucleonic matter are given in Eqs. (1) and (2)
in~\cite{ColucciSedrakian13}, respectively.

Our Set II consists of EOSs labeled ACB4-7. The low-density regime
($n\le n_0$) of these EOSs is equivalent to Set I.  Above the
saturation density $n_0=0.16$ fm$^{-3}$, but below the deconfinement
phase transition, EOSs ACB4 and ACB5 correspond to the stiffest EOS
of~\cite{Hebeler13}, while the EOSs ACB6 and ACB7 fit the
density-dependent relativistic mean field EOS
DD2-p30~\cite{Alvarez-Castillo:2016oln} accounting for nucleonic
excluded volume effects in that region.

In general, there exist two prescriptions for matching the low-density
nucleonic EOS to the quark matter EOS; which one is realized in nature
depends on the surface tension between nuclear and quark
matter~\cite{glenden_book,weber_book}.  If the tension between these
phases is low, a mixed phase of quark and nucleonic matter is formed
in-between purely nuclear and quark matter phases. Conversely, if the
tension is high, a sharp transition boundary is energetically
favorable. In the latter case, there is a jump in the energy density
at a certain transition pressure at which the baryochemical potentials
of both phases coincide. Since the surface tension is presently not
known accurately, both prescriptions are viable. Here we consider the
second case assuming that the surface tension between the quark matter
and nucleonic phases is high enough to sustain a sharp boundary
between them. In all our models the pressure matching between the
phases is performed via a standard Maxwell construction.

For the Set I quark matter EOS, we use the constant speed of sound
parametrization~\cite{AlfordCSS13}, see
also~\cite{Seidov1971,Zdunik13,AlfordSedrakian17}. The pressure beyond
the point where the phase transitions to quark matter takes place is
given analytically by
\bea\label{eq:CSS}
P(\ep) =\left\{
\begin{array}{ll}
P_{\rm tr}, & \ \ep_1\le \ep \le \ep_2,\\
P_{\rm tr}+ c_s^2(\ep-\ep_2), & \ \ep >\ep_2 ,
\end{array}
\right.
\eea
where $P_{\rm tr} = P(\ep_1) = P(\ep_2)$ is the value of the
(transition) pressure in the energy density range $\ep_1\le \ep \le
\ep_2$, and $c_s$ is the sound speed of the quark matter phase. It is
convenient to parametrize the magnitude of the jump via a parameter
$j$, as $\Delta\ep \equiv \ep_2-\ep_1 = \ep_{1}\, j$.  Within Set I we
consider two subsets that we call ``ACS-I'' and  ``ACS-II''. The values
of the parameters for these EOSs are presented in Table~\ref{param-acs}.

The ACS-I models, as we shall see below, generate high mass
$M/M_{\odot} \simeq 2$ twins, as well as twins when $j=0.6$. The
ACS-II
models produce low-mass $M/M_{\odot} \simeq 1.5$ HSs as well as twins
for $j=0.8$ and 1.0.
In the ACS-II models, the choice of
maximally stiff quark matter EOS allows for massive $\sim 2M_{\odot}$
compact stars. EOSs with these properties have been obtained recently
within a relativistic density functional approach to quark
matter~\cite{2012A&A...539A..16B,2013A&A...559A.118A,Kaltenborn:2017hus}.

\begin{table}[!htb]
	\centering
	\caption{Parameters for the ACS-I and ACS-II EOS models that
          adopt constant speed of sound parametrization. The
          parameters have the same meaning as in Eq. (1) in the main text,
          and $j$ parametrizes the phase transition energy
          density jump as $\Delta\ep \equiv \ep_{1}\, j$. The
          last column gives the maximum masses $M_{\rm
            max}$. \label{param-acs}}
	\begin{tabular}{c|c|ccc|c}
		\hline \hline
		&&$P_{\rm tr}$&$\epsilon_{\rm 1}/c^2$&$(c_s/c)^2$ &$M_{\rm max}$\\
		ACS&j&[$10^{34}{\rm dyn\ cm}^{-2}$] &[$10^{14}{\rm g\ cm}^{-3}$] & &$[M_\odot$]\\
                \hline
	      I  & 0.10   & 17.0 & 8.34 & 0.8 & 2.47  \\
	         & 0.27   & 17.0 & 8.34 & 0.8 & 2.31  \\
	         & 0.43   & 17.0 & 8.34 & 0.8 & 2.17  \\
	         & 0.60   & 17.0 & 8.34 & 0.8 & 2.05  \\
		\hline
	      II & 0.80   & 8.34 & 6.58 & 1.0 & 2.08  \\
	         & 1.00   & 8.34 & 6.58 & 1.0 & 1.97  \\
		\hline \hline
	\end{tabular}
\end{table}

For the Set II (labeled ``ACB'') we employ a piecewise polytropic
representation~\cite{Read2009,Hebeler13,Raithel:2016bux} of the EOS at
supersaturation densities ($n_1<n<n_5\gg n_0$)
\bea
\label{polytrope}
P(n) =  \kappa_i  (n/n_0)^{\Gamma_i}, \  n_i < n < n_{i+1}, \ i=1 \dots 4,
\eea 
where $\Gamma_i$ is the polytropic index in one of the density regions
labeled by $i=1\dots 4$.  The first polytrope describes a stiff
nucleonic EOS. The second polytrope corresponds to a first-order phase
transition with a constant pressure $P_{tr}=\kappa_2$
($\Gamma_2=0$). The polytropes in regions 3 and 4 above the phase
transition correspond to high-density matter, e.g., stiff quark
matter.

\begin{table}[!htb]
	\centering
	\caption{
          EOS models ACB4-ACB7.  The parameters have the same meaning
          as in Eq. (2) in the main text. The first polytrope ($i=1$)
          parametrizes the nuclear EOS at supersaturation densities,
          the second polytrope ($i=2$) corresponds to a first-order
          phase transition with a constant pressure $P_{tr}$ for
          densities between $n_2$ and $n_3$.  The polytropes in
          regions 3 and 4 above the phase transition correspond to
          high-density matter, e.g., quark matter.  The last column
          gives the maximum masses $M_{\rm max}$ on the hadronic
          (hybrid) branch corresponding to region 1 (4). The minimal
          mass $M_{\rm min}$ on the hybrid branch is given for region
          3.  }
	\label{param-123}
	\begin{tabular}{c|c|cccc|c}
		\hline \hline
		&&$\Gamma_i$&$\kappa_i$&$n_i$ &$m_{0,i}$&$M_{\rm max/min}$\\		
		ACB&i&&[MeV/fm$^3$]&[1/fm$^3$] &$[MeV]$&$[M_\odot$]\\		
\hline
		4&1&4.921&2.1680&0.1650&939.56 & 2.01  \\
		&2&0.0&63.178&0.3174&939.56 & -- \\
		&3&4.000&0.5075&0.5344&1031.2 & 1.96  \\
		&4&2.800&3.2401&0.7500&958.55 & 2.11  \\
		\hline
		5&1&4.777&2.1986&0.1650&939.56 & 1.40 \\
		&2&0.0&33.969&0.2838&939.56 & -- \\
		&3&4.000&0.4373&0.4750&995.03 & 1.39 \\
		&4&2.800&2.7919&0.7500&932.48 & 2.00 \\
		\hline
		6&1&4.2602&2.3096&0.1650&939.56& 2.00 \\
		&2&0.0&78.329&0.3659&939.56& -- \\
		&3&4.000&0.3472&0.6201&1050.3& 1.93 \\
		&4&2.800&2.7589&0.9000&964.49& 2.00 \\
		\hline
		7&1&4.408&2.2773&0.1650&939.56& 1.50 \\
		&2&0.0&41.316&0.3088&939.56& --\\
		&3&4.000&0.4124&0.5062&1003.20& 1.49 \\
		&4&2.800&4.9726&0.8300&883.29& 2.00 \\
		\hline \hline
	\end{tabular}
	\label{tab:1}
\end{table}

The Set II EOS parameters are given in Table~\ref{tab:1}. Note that
ACB5 (ACB7) requires that the phase transition onset occurs at a
nucleon number density of 0.284 (0.309) $\mathrm{fm}^{-3}$, i.e., at
roughly two times the nuclear saturation density.  While this density
seems to be on the low end, we recall that in our case we have
asymmetric nuclear matter, so that a transition at $2n_0$ is
equivalent to a transition density of $3-4 n_0$ for symmetric matter.
In particular, it is well known that isospin-symmetric systems are
bound in atomic nuclei whereas there are no bound systems of only
neutrons known in nature.  This is due to the effect of the asymmetry
energy in nuclear matter which stiffens isospin-asymmetric systems as
compared to the symmetric case. The consequence that the onset density
for deconfinement is lowered with increasing asymmetry has been
discussed in the literature before. For example, the effect is
illustrated in Fig. 1 of~\cite{DiToro:2009ig} for the example of
collisions of asymmetric nuclei (Au+Au), where a lowering of the onset
density to $4n_0$ from $\sim 6 n_0$ in the symmetric case was
obtained. For beta-equilibrated matter in neutron stars the lowering
is still more pronounced. In Fig. 13 of~\cite{Fischer:2017zcr}, the
phase transition onset density of $2.5 n_0$ has been obtained for a
model of beta-equilibrated matter with a Maxwell construction, while
effects of inhomogeneities may lower the onset density further. For
the extreme case of a Glendenning construction~\cite{Glendenning92}
(neglecting surface tension), the lowering of the phase transition
density has been shown in Fig. 1 of~\cite{Sagert:2008ka}. A possible
onset density of $2 n_0$ in neutron star matter is realistic and does
not contradict the phenomenology of heavy-ion collisions, where in
almost symmetric nuclear matter the transition is not expected at
these low densities.

\begin{figure}[htb]
 \includegraphics[width=8.5cm]{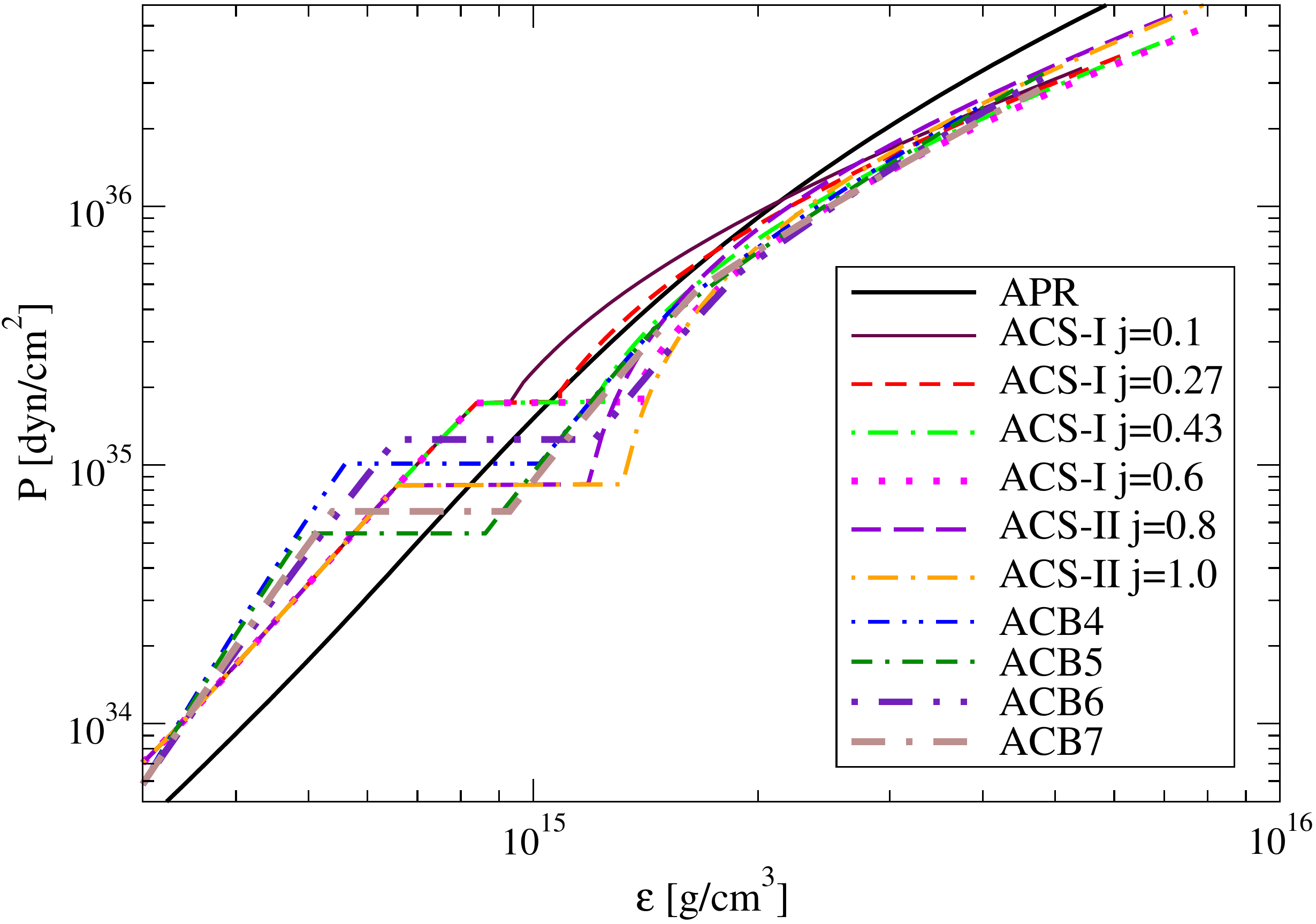}
 \caption{Pressure vs energy density for the equations of state with
   first-order phase transitions we consider in this work.}
 \label{fig:Pvsrho}
 \end{figure}

 All models have been supplemented with the low-density EOSs of crustal
matter according to~\cite{BPS1971,NV1973}. The pressure vs energy
density for each EOS in our sample is plotted in Fig.~\ref{fig:Pvsrho}.

\section{Methods}

To test whether the Set I and Set II EOSs satisfy
the TD constraints set by GW170817, we adopt tabulated versions of the
EOSs we developed and compute sequences of nonrotating HSs for each
EOS. For every member of the sequence, we compute the dimensionless TD
parameter $\Lambda = \lambda^{\rm (tid)}/M^5$ (in the small tidal
deformation approximation), where $\lambda^{\rm (tid)}$ is the stellar
TD parameter and $M$ the stellar gravitational mass. For more details
on the calculation of $\Lambda$,
see~\cite{Hinderer:2007mb,Damour:2009vw,Binnington:2009bb,Yagi:2013awao}.

To investigate the I-Love-Q relations for both slowly and rapidly
rotating HSs we compute the dimensionless moment of inertia $\bar
I=I/M^3$ (with $I$ being the stellar moment of inertia), $\Lambda$,
and the dimensionless quadrupole moment $\bar Q=-Q/(M^3\chi^2)$. Here,
$Q$ is the spin-induced quadrupole moment
(see~\cite{FSbook,Pappas:2012ns}), and $\chi=J/M^2$, with $J$ the
total angular momentum. The calculation of these quantities in the
slow-rotation approximation is performed as in~\cite{Yagi:2013awao}
following the Hartle-Thorne
formalism~\cite{Hartle:1967he,Hartle:1968si}.
For rapidly rotating stars, we compute $\bar I$ and $\bar Q$ for
sequences of self-consistent, rotating stellar equilibrium
configurations that we build with the code
of~\cite{CST92,Cook94,CST94b,CST94a}. More specifically, $\bar Q$ is
calculated through the asymptotic structure of the spacetime as
described in~\cite{Doneva:2013rha}.  We checked the consistency of the
two different codes in the slow-rotation regime.


\begin{figure}[htb]
\includegraphics[width=8.5cm]{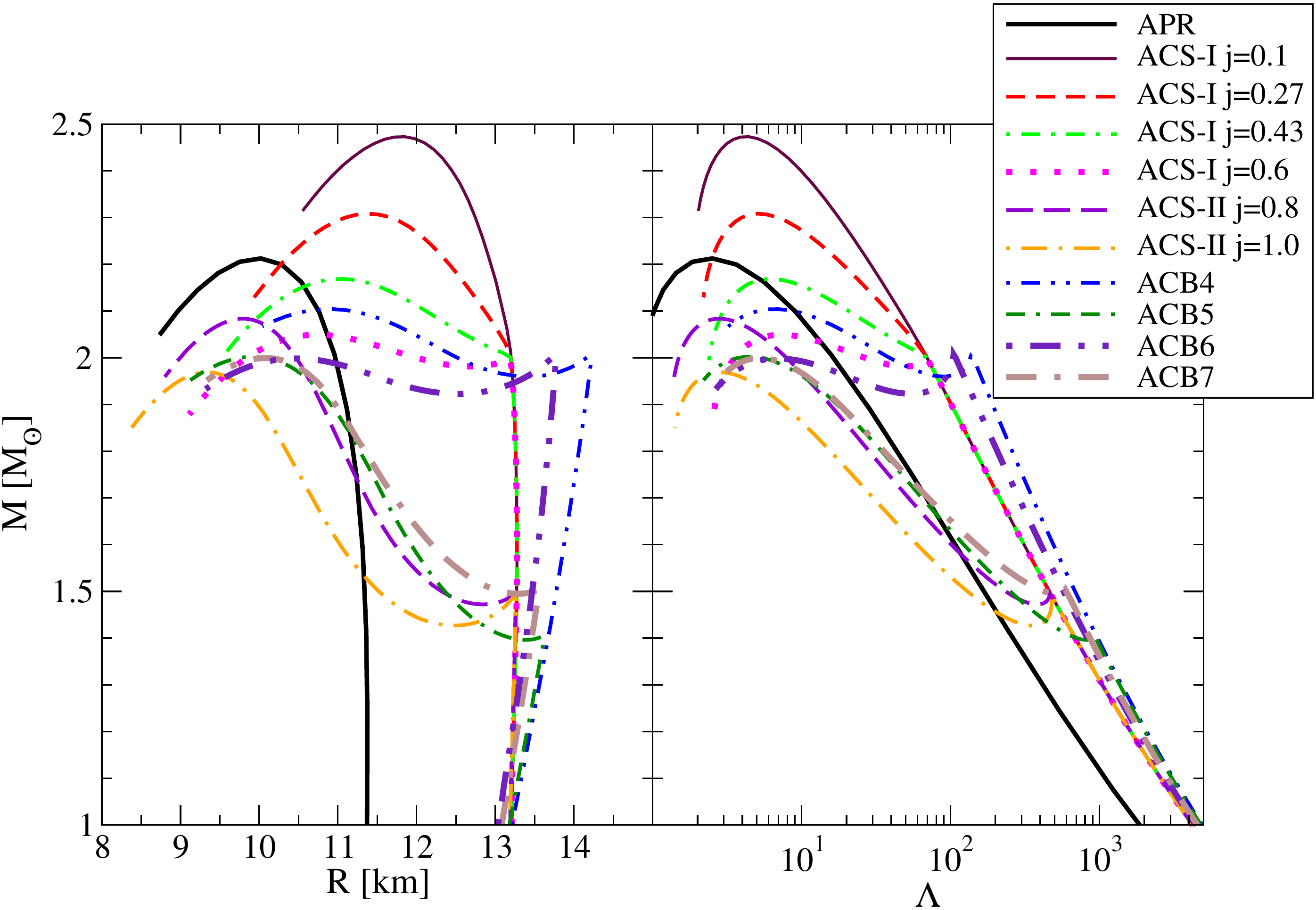}
\caption{Left: Mass-radius relation for nonrotating HSs. Observe that
  some EOSs admit a stable third family branch (separate from the
  stable NS branch). Right: Mass-TD relation for nonrotating stars
  with the same EOSs. The APR EOS is also shown for comparison.}
\label{fig:MR}
\end{figure}

\section{Results}

The $M-R$ (here $R$ is the stellar areal radius)
relations for nonrotating HSs with the Set I and Set II EOSs are
shown in the left panel of Fig.~\ref{fig:MR}. For comparison we also
show the APR EOS~\cite{PhysRevC.58.1804}. The $M-R$ plot demonstrates
that for the ACS-I $j=0.43$, ACS-I $j=0.6$, ACS-II and ACB EOSs a
third family of hybrid hadron-quark stars emerges. All EOS
parametrizations we developed here satisfy the $2M_\odot$ bound for
the maximum mass. The ACS-I $j=0.6$, ACB 4 and ACB 6 models give rise
to high mass ($\sim 2.0M_\odot$) twins, while the ACS-II and ACB 5 and
ACB 7 models lead to low-mass ($\sim 1.5M_\odot$) HSs.

The right panel of Fig.~\ref{fig:MR} shows the $M-\Lambda$ relation
for the same configurations as in the left panel. Notice that when the
third family branch emerges at low masses, $\Lambda(M)$ can no longer
be approximated as a linear function as was found
in~\cite{PhysRevD.85.123007}.  As a result, the method adopting this
approximation to estimate the tidal deformability of a $1.4M_\odot$
compact object~\cite{PhysRevLett.111.071101}, and which was used in
the case of GW170817~\cite{TheLIGOScientific:2017qsa}, excludes the
possibility of low-mass HSs in the third family.


\begin{figure*}[htb]
  \includegraphics[width=8.5cm]{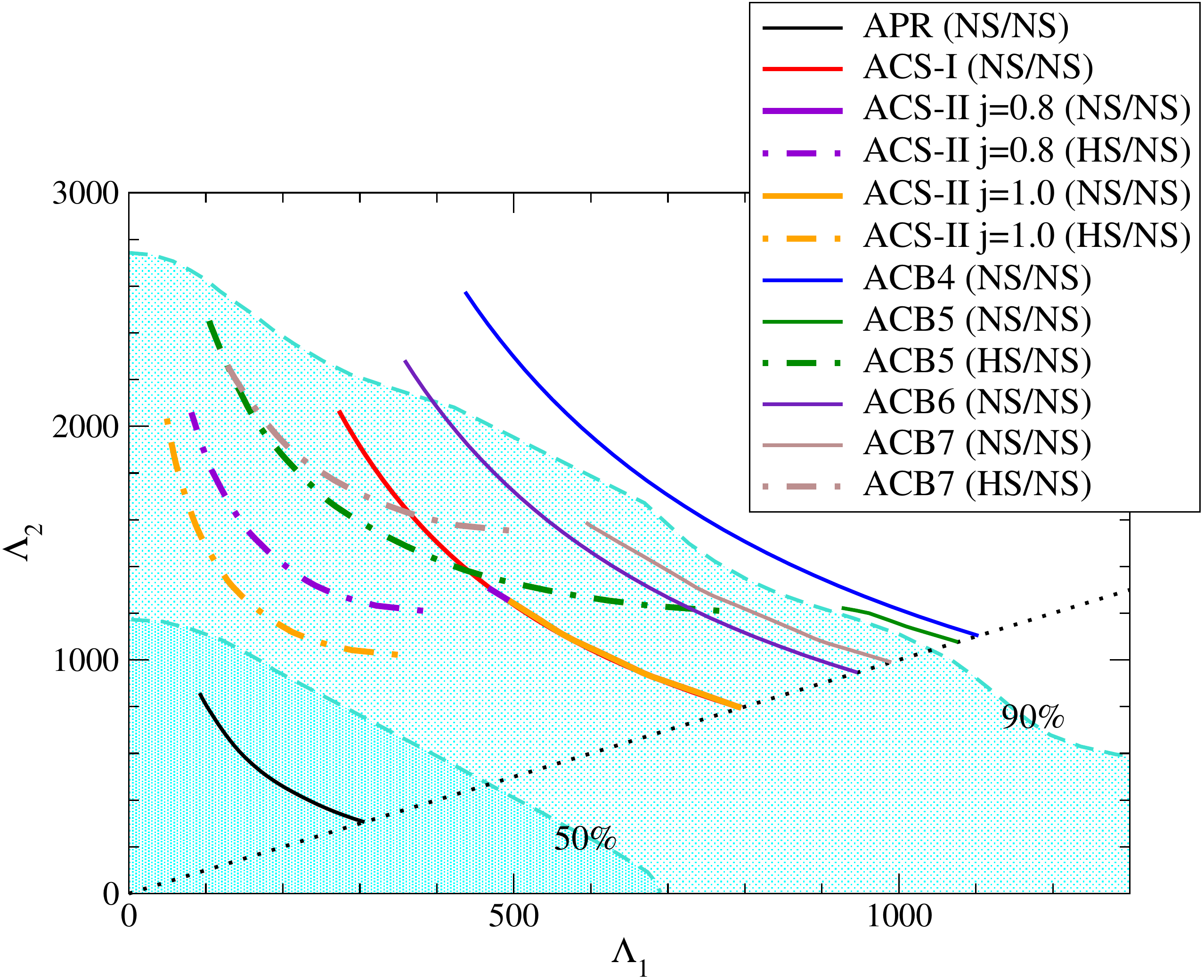}
  \includegraphics[width=8.5cm]{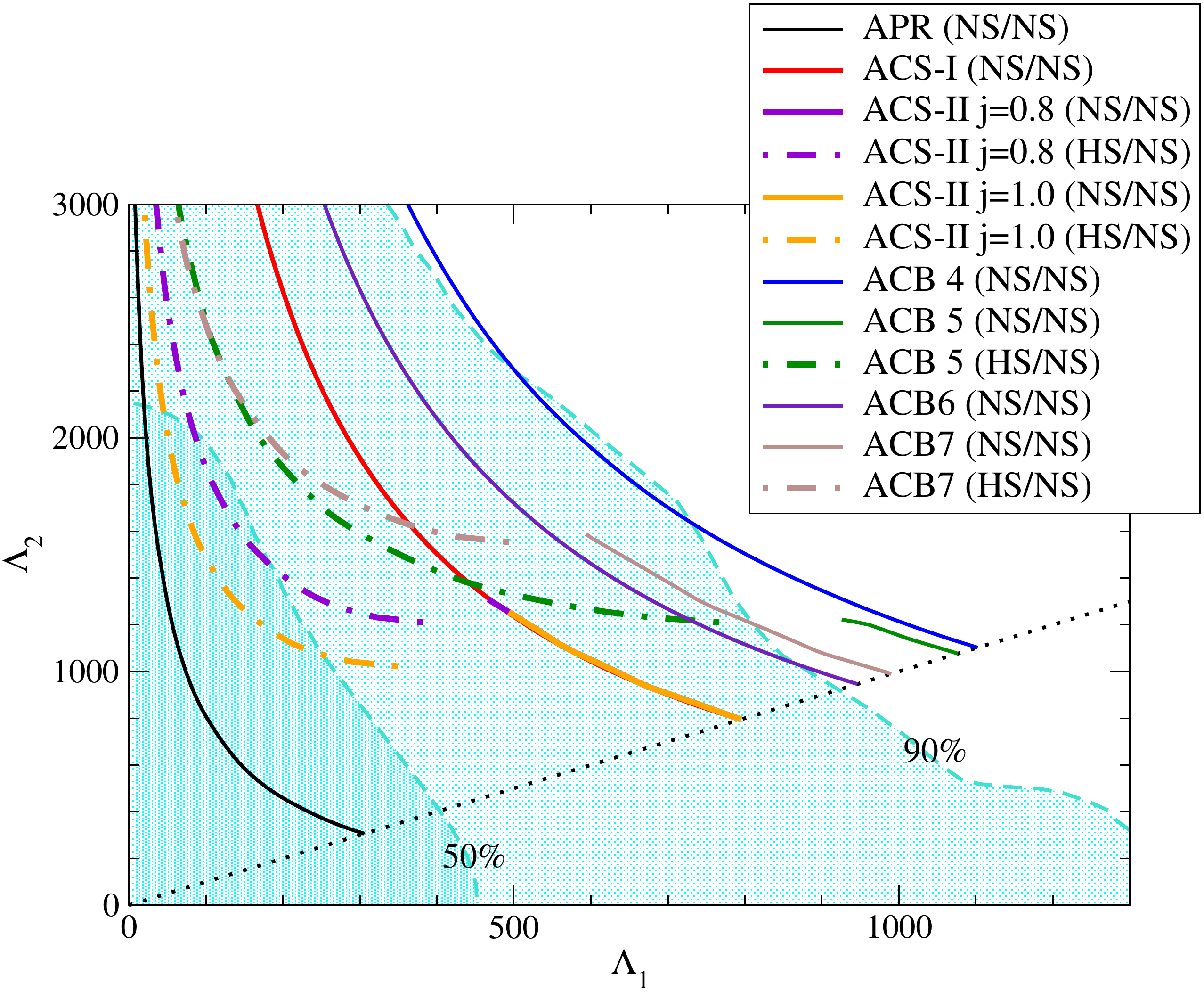}
\caption{TDs of compact objects in GW170817 with chirp mass
  $\mathcal{M} = 1.188 M_\odot$ for the ACS-I (which are
  indistinguishable from each other for $\Lambda \geq 70$), ACS-II and
  ACB EOSs. Solid curves correspond to both stars being in the NS
  branch while dashed-dotted curves correspond to one of the stars
  being in the third family (namely HSs). Note that the gap between
  the solid and dashed-dotted component of a given color curve arises
  because the HS member in the binary is unstable.  Only the plot
  above the black dotted line ($\Lambda_1 = \Lambda_2$) is
  relevant. The dark and light cyan shaded areas correspond to the
  parameter region within the 50\% and 90\% credible upper bound set
  by GW170817 with prior $|\chi| < 0.05$ (left panel) and $|\chi| <
  0.89$ (right panel. The solid black curve corresponds to the
  APR EOS. }
\label{fig:Lambda12}
\end{figure*}

In Fig.~\ref{fig:Lambda12} we plot the TDs $\Lambda_1$ vs $\Lambda_2$
for binary compact objects with the chirp mass of GW170817, i.e.,
$\mathcal{M} = 1.188 M_\odot$. We show curves corresponding to Set I
and Set II EOSs, as well as the 50\% and 90\% credible constraints set
by GW170817 with low-spin prior $|\chi| < 0.05$ (in the left
panel). We find that the ACS-I and ACS-II, ACB5-7 EOSs are consistent
with the 90\% credible upper bound set by GW170817, while ACB4 is
inconsistent. Notice that different ACS-I EOSs are indistinguishable
from each other for $\Lambda \geq 70$ (see the right panel of
Fig.~\ref{fig:MR}). Thus, we only show one curve for ACS-I.  An
important finding is that while certain hadronic EOSs may not satisfy
the GW170817 constraints on the TD, they can become compatible with
GW170817 if a first-order phase transition occurs in one of the stars
and it is a HS in the 3rd family branch. This is exemplified by our
ACB4 and 5 models which are matched to the same hadronic baseline
EOS. In particular, the hadronic (solid) parts of the
$\Lambda_1-\Lambda_2$  curve that corresponds to them are excluded by
GW170817 at 90\% confidence, but the HS-NS (dashed-dotted) ACB5 curve
satisfies the GW170817 90\% confidence constraints on the TD. Apart
from ACB5, the ACB7 and ACS-II EOSs satisfy the constraints set by
GW170817 when having one star in the normal hadronic branch and the
other one in the third family (the dot-dashed branch of a given color
curve). Therefore, \emph{GW170817 is consistent with the coalescence
  of a binary HS-NS.}
  
One finds similar results for the high-spin prior GW170817 TD
constraints within the range $\Lambda_2 < 3000$ (see right panel in
Fig.~\ref{fig:Lambda12}), but ACB4 is only marginally inconsistent with
the 90\% credible constraints. However, since the mass of the primary star
can be as large as $2.26M_\odot$, the HS/NS scenario may be consistent
with GW170817 even for equations of state with the transition
happening in the high mass regime. Investigating this possibility
further is beyond the scope of this paper as the bound in the
$\Lambda_1$-$\Lambda_2$ plane for $\Lambda_2>3000$ is not provided
in~\cite{TheLIGOScientific:2017qsa}.

\begin{figure*}[htb]
\includegraphics[height=7.cm]{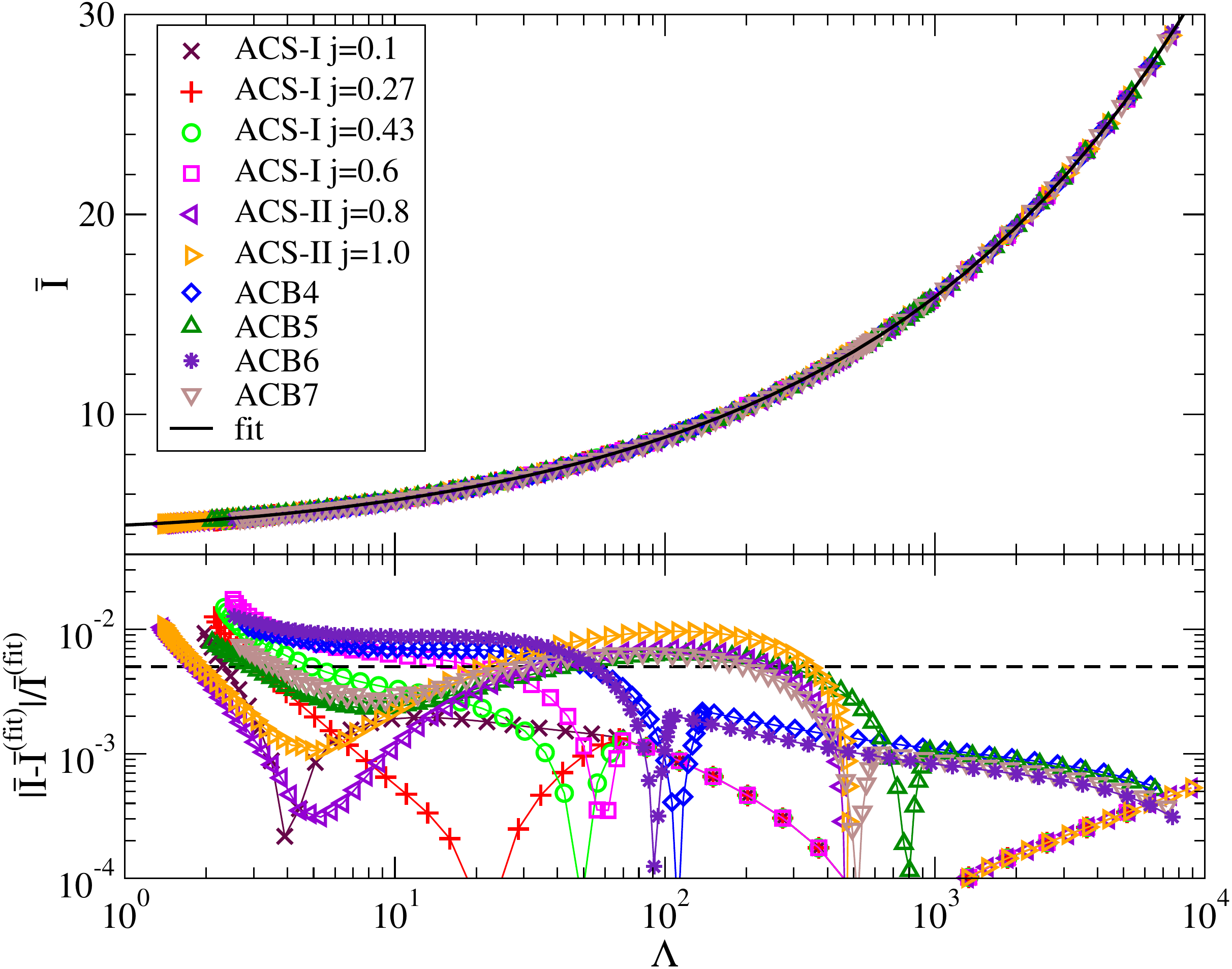}
\includegraphics[height=7.cm]{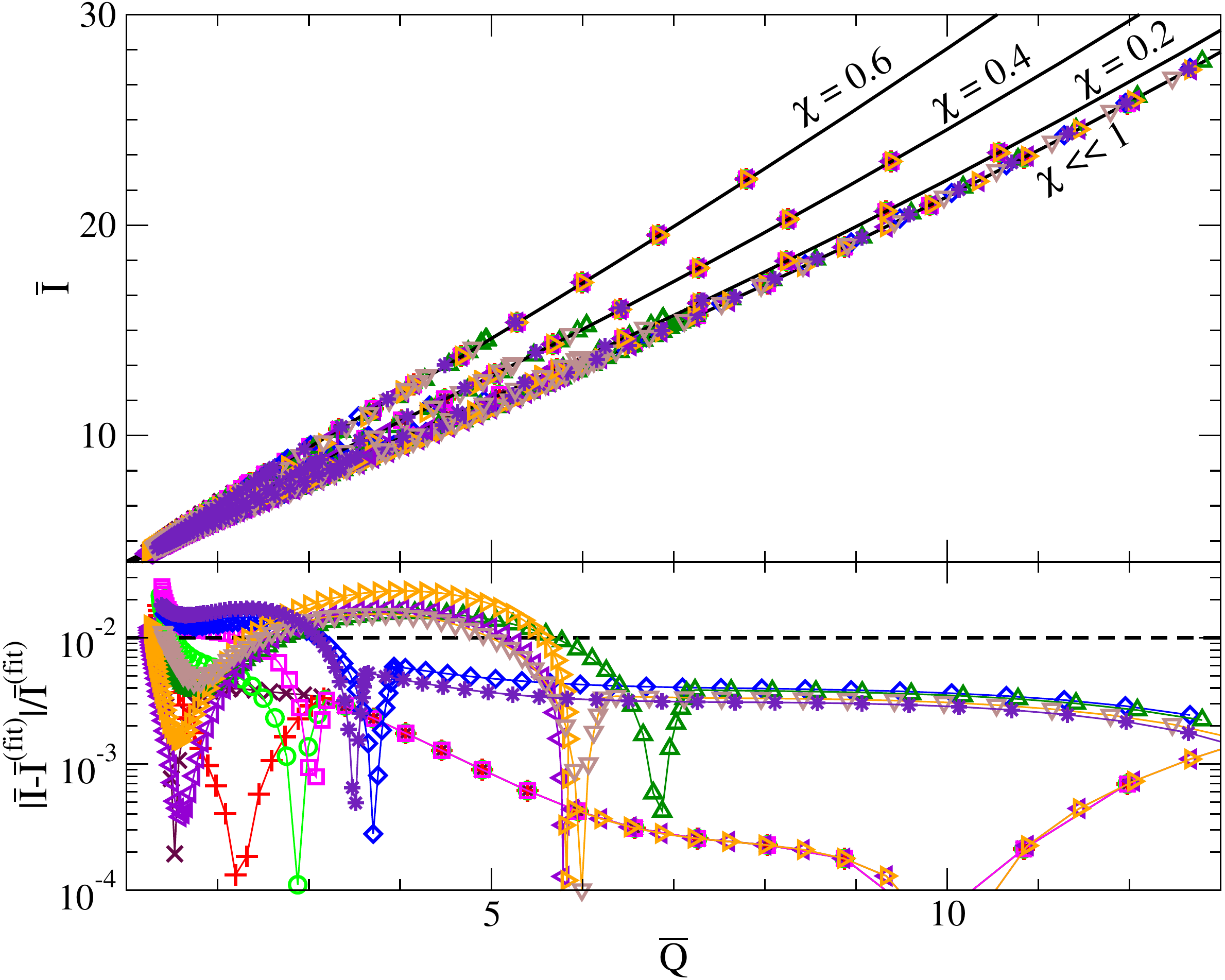}
\caption{Left: I-Love relations for slowly rotating stars together with
  the analytic fit of~\cite{Yagi:2016bkt}. Right: I-Q relation for
  slowly and rapidly rotating stars with various dimensionless spin
  parameters $\chi$ together with the analytic fit
  in~\cite{Chakrabarti:2013tca}. In both panels $ |\bar {\rm I}-\bar
  {\rm I}^{\rm (fit)}|/{\rm I}^{\rm (fit)}$ is the fractional
  difference between the numerical results and the corresponding
  analytic fits. The horizontal dashed lines in the $ |\bar {\rm
    I}-\bar {\rm I}^{\rm (fit)}|/{\rm I}^{\rm (fit)}$ plots indicate
  the maximum fractional difference between the numerical results and
  the fits reported in~\cite{Yagi:2016bkt}.}
\label{fig:I-Love-Q}
\end{figure*}

In Fig.~\ref{fig:I-Love-Q} we show the I-Love-Q relations for slowly
rotating and rapidly rotating stars constructed with the Set I and Set
II EOSs. Following~\cite{Chakrabarti:2013tca} we generated
$\chi$-constant sequences for rapidly rotating stars. Notice that the
relations for slowly rotating stars remain universal and agree with
those reported for neutron stars and quark stars
in~\cite{Yagi:2016bkt}. Notice also that the I-Q relations for rapidly
rotating stars remain universal for a fixed $\chi$, in agreement
with~\cite{Chakrabarti:2013tca,Pappas:2013naa}. The deviations from
universality with the hybrid hadron-quark EOSs considered here are
slightly larger than those in~\cite{Yagi:2016bkt}, especially for high
mass (small $\Lambda$ or $\bar Q$) stars. Nevertheless, the relations
remain universal within $\sim 3\%$ for both slowly and
rapidly rotating stars. Thus, our results extend the previously
discovered universal I-Love-Q relations for compact stars into the
third family. 

\section{Discussion}

We have constructed hybrid hadron-quark EOSs
that: i) give rise to a third family of compact objects, ii) are
consistent with the existence of $2M_\odot$ pulsars, and iii) result
in low-mass twins ($\sim 1.5M_\odot$). Using our new model EOSs we
computed the TD of sequences of relativistic stars. In contrast to
realistic neutron star EOSs, where the dimensionless tidal
deformability can be approximated as a linear function of the
gravitational mass in the vicinity of $1.4M_\odot$, in the case of
hybrid hadron-quark EOSs with low-mass twins this is no longer true. As
a result, using this approximation to estimate the tidal deformability
of a $1.4M_\odot$ compact objects should be avoided because it
excludes the possibility of testing for HSs. All EOSs in our sample,
except for one, are consistent with the GW170817 90\% confidence TD
bounds. We discover that while a sufficiently stiff hadronic baseline
EOS may be inconsistent with GW170817, a hadron-quark phase transition
in the compact object interior can soften the EOS to make it
compatible with GW170817. Importantly, we find that GW170817 is
entirely consistent with coalescence of a binary hybrid hadron-quark
-- neutron star.

Furthermore, we computed the I-Q relations~\footnote{We do not compute
  the I-Love relations for rapidly rotating relativistic stars as the
  formulation for computing the tidal Love numbers for such stars is
  currently lacking. Various tidal Love numbers for slowly rotating
  stars have been computed to leading order in spin
  in~\cite{Pani:2015nua}, though the spin correction to $\Lambda$ that
  we consider here enters first at quadratic order in spin.}  for
rotating relativistic stars adopting our new hybrid hadron-quark EOSs,
and discover that despite the sharp first-order phase transition at
the hadron-quark interface in the interior of these stars, the hybrid
star I-Q relations agree with the I-Q relations of slowly and rapidly
rotating realistic neutron stars and quark stars to better than $\sim
3\%$. Therefore, the I-Love-Q relations can be adopted to either
perform equation-of-state independent tests of general relativity or
to break degeneracies in parameter estimation from GWs even when HSs
in the third family are present.

Future GW observations will help understand the properties of hybrid
stars and resolve the current controversy about the nature of the
hadron-to-quark matter transition at zero temperature: is it a
first-order transition with large jump in energy density or is it a
smooth crossover? At this time, it seems that the only possible way to
constrain the nature of binary compact objects through GWs, and hence
to resolve the aforementioned controversy requires GW detectors that
are sensitive in the high frequency regime, where tidal effects are
strong and can lead to measurable deviations between the GWs generated
by binary NS-NS and binary HS-NS. To address this point theoretically
it is necessary to perform binary HS-NS simulations in full general
relativity and compare them to NS-NS simulations. Our work sets the
foundations for performing such an analysis by constructing the equations
of state that respect all currently known constraints.

Another way to probe the aforementioned controversy is to combine GW
and electromagnetic observations of compact objects. For example, if
the presently ongoing NICER~\cite{doi:10.1117/12.2231304} measures the
radius of the $1.44M_\odot \pm 0.07$ pulsar J0437-4715 to not be less
than 14 km with an uncertainty of less than 500 m, then soft hadronic
EOSs would be incompatible, and the stiff hadronic baseline of the set
of EOSs discussed here would be favored. However, according
to~\cite{Annala:2017llu}, a hadronic EOS with $R_{1.4} > 13.4$ km is
inconsistent with GW170817; thus, the HS-NS scenario for GW170817 would
be a most likely explanation, implying an EOS with a stiff hadronic
part and a strong phase transition.

\section*{Acknowledgments}
D.E.A-C. is grateful for support by the ExtreMe Matter Institute EMMI
at the GSI Helmholtzzentrum f\"ur Schwerionenphysik Darmstadt,
Germany, as well as to the Heisenberg-Landau and Bogoliubov-Infeld
programs for collaboration between JINR Dubna and Institutions in
Germany and Poland, respectively.  D.B. is supported by the Russian
Science Foundation under Contract No. 17-12-01427. D.E.A-C.,
D.B. and A.S.  acknowledge partial support by the COST Action MP1304
(NewCompStar) for networking activities. A.S. is supported by the
Deutsche Forschungsgemeinschaft (Grant No. SE 1836/3-2). K.Y. would
like to acknowledge networking support by the COST Action GWverse
CA16104.

\bibliography{ref}
\end{document}